\documentclass[letterpaper,american,prl, showpacs, twocolumn]{revtex4}

\usepackage[T1]{fontenc}
\usepackage[latin1]{inputenc}
\usepackage{amsmath}
\usepackage{babel}
\usepackage{graphics}
\usepackage{amssymb}

\makeatletter

\providecommand{\LyX}{L\kern-.1667em\lower.25em\hbox{Y}\kern-.125emX\@}

\makeatother
\begin{document}

\newcommand{\epshat}{\hat{\varepsilon }}

\newcommand{\eps}{\varepsilon }

\title{Prevalence of unstable attractors in networks of pulse-coupled oscillators }

\author{Marc Timme}

\author{Fred Wolf }

\author{Theo Geisel}

\affiliation{Max-Planck-Institut für Strömungsforschung and Fakultät für Physik,
Universität Göttingen, 37073 Göttingen, Germany.}

\begin{abstract}
We present and analyze the first example of a dynamical system that
naturally exhibits attracting periodic orbits that are \textit{unstable}.
These unstable attractors occur in networks of pulse-coupled oscillators,
and become prevalent with increasing network size for a wide range
of parameters. They are enclosed by basins of attraction of other
attractors but are remote from their own basin volume such that arbitrarily
small noise leads to a switching among attractors. 
\end{abstract}


\maketitle
As attractors determine the long-term behavior of dissipative dynamical
systems, the concept of attractors is central to the analysis of many
natural systems as well as to the design of artificial systems. For
instance, the computational capabilities of neural networks are controlled
by the attractors of their collective dynamics. Consequently, the
nature and design of attractors in such systems constitute a focus
of current research \cite{Mirollo,Ernst,Tsodyks_et_al}. In general,
the state space of a nonlinear dynamical system is partitioned into
various basins of attraction from which states evolve towards the
respective attractors. Since states that are slightly perturbed from
an attractor often stay confined to its vicinity and eventually return
to the attractor, attractors are commonly considered to be \textit{stable}
\cite{Eckmann_et_al}. 

In the present letter, we show that \textit{unstable} attractors exist
and arise naturally as a \textit{collective} phenomenon in networks
of pulse-coupled oscillators \cite{Mirollo,Ernst}, which where introduced
to model e.g.\ synchronization in spiking neural networks and the
dynamics of other natural systems such diverse as cardiac pacemaker
cells, populations of flashing fireflies, and earthquakes (cf.\ \cite{Mirollo,Ernst,Tsodyks_et_al,Peskin_et_al}).
We identify an analytically tractable network exhibiting unstable
attractors. For this network we demonstrate the existence of attractors
that are linearly unstable and are thus separated from the volume
of their own basins of attraction. Such attracting yet unstable states
are consistent with a definition of attractors introduced by Milnor,
which neither presumes nor implies stability \cite{Milnor_et_al}.
In some other systems such Milnor attractors might not be uncommon
if they are \textit{strange} attractors that display irregular dynamics
\cite{Sommerer_et_al}. More generally, however, attractors that are
not stable seem to be special cases that have to be constructed artificially
by precisely tuning parameters. Contrary to this intuition, we report
here that unstable attractors with regular, \textit{periodic} dynamics
are typical in large networks and persist even if the physical parameters
are varied substantially. 

We argue that dynamical consequences of unstable attractors may persist
in a general class of systems of pulse-coupled units. Such consequences
include an ongoing switching among unstable attractors in the presence
of noise. In systems where the convergence towards an attractor has
a functional role, such as the solution of a computational task by
a neural network \cite{Hopfield_et_al}, switching induces a high
degree of flexibility that provides the system with a unique advantage
compared to multistable systems: It will be hard to leave a stable
attractor after convergence, e.g.\ the completion of a task. With
an unstable attractor, however, a small perturbation is sufficient
to leave the attractor and to switch towards another one. 

We consider a homogeneous network of \( N \) all-to-all pulse-coupled
oscillators with delayed interactions. A phase variable \( \phi _{i}(t)\in [0,1] \)
specifies the state of each oscillator \( i \) at time \( t \).
Its free dynamics is given by \begin{equation}
\label{eq:free_dynamics}
d\phi _{i}/dt=1.
\end{equation}
Whenever oscillator \( i \) reaches a threshold, \( \phi _{i}(t)=1 \),
the phase is reset to zero, \( \phi _{i}(t^{+})=0 \), and a pulse
is sent to all other oscillators \( j\neq i \), which receive this
signal after a delay time \( \tau  \). Depending on whether the input
\( \epshat  \) is subthreshold or suprathreshold this induces a phase
jump according to \begin{equation}
\label{eq:reception}
\phi _{j}((t+\tau )^{+})=\min \{U^{-1}(U(\phi _{j}(t+\tau ))+\hat{\varepsilon }),\, 1\}
\end{equation}
which depends on the phase \( \phi _{j}(t+\tau ) \) of the receiving
oscillator and the effective excitatory coupling \( \epshat =\eps /(N-1)>0 \).
The function \( U(\phi ) \) is twice continuously differentiable,
monotonously increasing, \( U'>0 \), concave (down), \( U''<0 \),
and normalized such that \( U(0)=0 \), \( U(1)=1 \). For many models
of biological systems \( U(\phi ) \) represents a 'potential' of
an oscillator at phase \( \phi  \). For a more detailed discussion
of the model see references \cite{Mirollo,Ernst}.

For such pulse-coupled systems, periodic orbits with groups of synchronized
units constitute relevant attractors \cite{Mirollo,Ernst,Tsodyks_et_al,Peskin_et_al}.
For instance, the network described above possesses a single global
attractor in which all oscillators are synchronized with zero phase
lag if the interactions are instantaneous (\( \tau =0 \)) \cite{Mirollo}.
Here we consider the case of delayed interactions (\( \tau >0 \))
where multiple different cluster-state attractors with several synchronized
groups of oscillators (clusters) coexist \cite{Ernst}. Such attractors
are period-one orbits with all oscillators in the clusters reaching
threshold and sending out pulses exactly once during each period. 
\begin{figure*}
\includegraphics{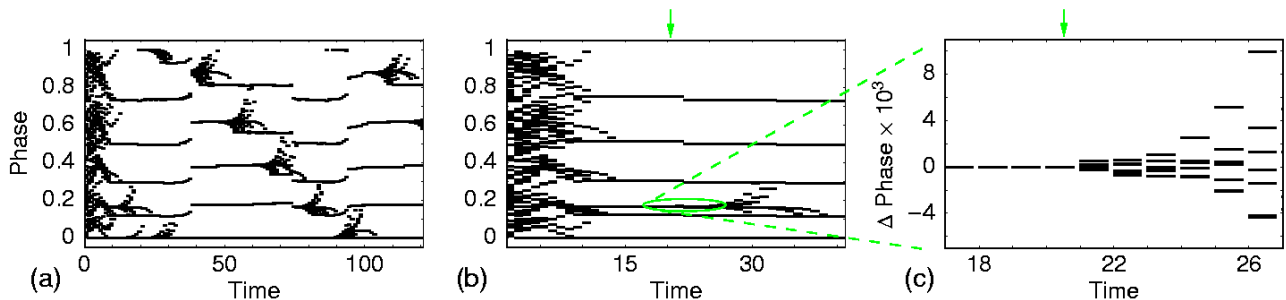}

\caption{Phase dynamics of a large network (\protect\( N=100\protect \),
\protect\( \eps =0.2\protect \), \protect\( \tau =0.15\protect \)).
Phases of all oscillators are plotted whenever a reference oscillator
has been reset. (a) Dynamics with noise (\protect\( \eta =10^{-3})\protect \),
(b) deterministic dynamics in response to a single phase perturbation
(arrow, \protect\( \sigma =10^{-3}\protect \)), note that the system
switches from a six-cluster to a five-cluster state, (c) phase differences
from the average phase of one cluster in response to the perturbation. }
\end{figure*}
We find (cf.\ Fig.\ 1a) that, although the system converges towards
a periodic orbit from random initial conditions, weak noise is often
sufficient to drive the system away from that attractor such that
successive switching towards different attractors occurs. This alternating
synchronization and desynchronization might be due to stable attractors
located close to the boundaries of their basins of attraction, such
that the noise drives the state of the system into a neighboring basin.
In an otherwise noiseless system we tested this possibility by applying
instantaneous perturbations of gradually decreasing strengths (down
to \( \sigma =10^{-8} \), cf.\ Fig.\ 1b,c). As we did not find a
strength for which \textit{any} of the perturbed states returned to
the attractor, we hypothesized that the persistent switching dynamics
(Fig.\ 1a) is due to attractors that are unstable. 

In order to verify this hypothesis directly, we analyze a small network
of \( N=6 \) oscillators for which instantaneous perturbations lead
to a similar switching among attractors. At given parameters \cite{parameters}
this network exhibits a set of period-one orbits that are related
by a permutation of phases in such a way, that the system may switch
among them (Fig.\ 2a, states on the periodic orbits marked in red,
yellow, blue). Due to their permutation-equivalence these orbits have
identical stability properties. The state of the network at time \( t \)
is specified by \( \boldsymbol {\phi }(t)=(\phi _{1}(t),\ldots ,\phi _{6}(t))^{\mathsf{T}} \),
such that the orbit marked in yellow in Fig.\ 2a is defined by the
initial condition \cite{ABC}\begin{equation}
\label{eq:syn}
\boldsymbol {\phi }(0)=(0,\, 0,\, A,\, A,\, B,\, C)^{\mathsf{T}}.
\end{equation}
 Here the origin of time was chosen such that oscillators \( 1 \)
and \( 2 \) have just sent a signal and have been reset. Moreover,
at \( t=0 \) only these two signals (and no others) have been sent
but not yet received. The numerical values for the particular parameters
considered, \( A\approx 0.176 \), \( B\approx 0.499, \) \( C\approx 0.747 \),
can be identified in Fig.\ 2a (orbit marked in yellow). This orbit
indeed is periodic, \( \boldsymbol {\phi }(T)=\boldsymbol {\phi }(0) \),
such that after the period \( T \) each oscillator has reached threshold,
has sent a signal and has been reset exactly once (for details see
\cite{Timme_inprep}).
\begin{figure}
{\centering {\small \resizebox*{75mm}{!}{\includegraphics{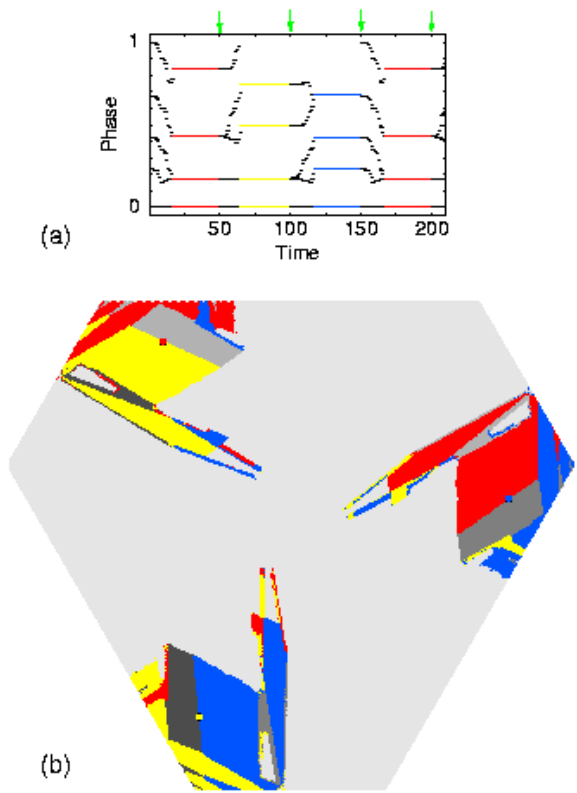}} } \par}

\caption{Small network (\protect\( N=6\protect \), \protect\( \eps =0.2\protect \),
\protect\( \tau =0.15\protect \)): (a) Noise-free phase dynamics
in response to single perturbations (arrows), (b) Basin structure
in a two-dimensional planar section through six-dimensional state
space. Small red, yellow, and blue disks represent points on the attractors
color-marked in (a). Their basins of attraction are marked in the
same colors. Medium gray areas are basins of permutation-related attractors,
lightest gray marks the union of the basins of all other attractors. }
\end{figure}

To perform a stability analysis, we define a return map by choosing
oscillator \( i=1 \) as a reference: Let \( \phi _{n,i}:=\phi _{i}(t_{n}) \)
be the perturbed phases of the oscillators \( i \) at times \( t_{n}>0 \),
\( n\in \mathbb {N} \), just after the resets of oscillator \( 1 \),
\( \phi _{1}(t_{n})\equiv 0 \). Thus the five-dimensional vector
\( \boldsymbol {\delta _{n}}=\boldsymbol {\phi _{n}}-(0,\, A,\, A,\, B,\, C')^{\mathsf{T}} \)
(see \cite{modeldetails}) defines the perturbations \( \delta _{n,i} \)
for \( i\in \{2,\ldots ,6\} \) where we choose \( 0<\delta _{n,2} \)
and \( \delta _{n,3}<\delta _{n,4} \) . Following the dynamics, the
five-dimensional return map is given by \cite{F_i} \begin{equation}
\label{eq:Poincare}
\boldsymbol {\delta }_{n+1}=\boldsymbol {F}(\boldsymbol {\delta }_{n}).
\end{equation}
 The linearized dynamics of a slightly perturbed state with split-up
clusters is described by the Jacobian matrix \( M=\left. \partial \boldsymbol {F}(\boldsymbol {\delta })/\partial \boldsymbol {\delta }\right| _{\boldsymbol {\delta }=\mathbf{0}} \).
It has four zero eigenvalues \begin{equation}
\label{eq:lambda_zero}
\lambda _{i}=0\, \, \, \mathrm{for}\, \, \, i\in \{1,2,3,4\}
\end{equation}
 such that a six-dimensional state-space volume accessed by the perturbation
is contracted onto a two-dimensional manifold. This reflects the fact
that suprathreshold input received simultaneously by two or more oscillators
leads to a simultaneous reset and thus a synchronization of these
oscillators independent of their precise phases. If a single oscillator
is reset by a suprathreshold input signal, it exhibits a precise lag
in firing time \( \Delta t=\tau  \) compared to the oscillator that
has sent this signal. In contradistinction, the concavity of \( U \)
implies that simultaneous subthreshold input to two or more oscillators
leads to an increase of their phase differences, i.e.\ a desynchronization
of oscillators with similar phases. For the orbits considered here,
this is reflected by the only non-zero eigenvalue\begin{equation}
\label{eq:positive_eigenvalue}
\lambda _{5}=\frac{(2U'(c_{0})-U'(a_{1}))U'(c_{1})U'(c_{2})U'(c_{3})}{U'(a_{1})U'(a_{2})U'(a_{3})U'(a_{4})}>1
\end{equation}
where \( c_{i}=\tau +a_{i} \) for all \( i \) (cf.\ \cite{ABC}).
Because \( c_{i}>a_{i}>c_{i-1} \) for all \( i \) and \( U'>0 \),
\( U''<0 \), this eigenvalue is larger than one, i.e.\ the periodic
orbit is linearly \textit{unstable}. If there is no homoclinic connection,
this implies that such an attractor is not surrounded by a positive
volume of its own basin of attraction, but is located at a distance
from it: Thus, every random perturbation to such an attractor state
-- no matter how small -- leads to a switching towards a different
attractor. Furthermore, this periodic orbit indeed is an \textit{attractor:}
Right after the perturbation off a periodic orbit (e.g. the one marked
in red in Fig.\ 2a, which is permutation-equivalent to the yellow
one) the state of the system is mapped onto a two-dimensional manifold,
re-synchronizing one cluster. The state then evolves towards a neighborhood
of another attractor (here: the yellow one) in a lower dimensional
effective state space without further dimensional reduction. Here,
forming the second cluster, suprathreshold input leads to the last
dimensional reduction while the state is mapped directly onto the
periodic orbit.

In general, a periodic orbit is \textit{unstable,} if after a random
perturbation into its vicinity, one or more clusters are not re-synchronized
by simultaneous suprathreshold input but desynchronize due to simultaneous
subthreshold input. An unstable \textit{attractor} results if these
clusters are formed through synchronization in a region of state space
that is \textit{separated} from the periodic orbit towards which the
state then converges. Roughly, unstable attractors can be viewed as
saddle periodic orbits together with a funnel mechanism that puts
trajectories onto its stable manifold (for details see \cite{Timme_inprep}). 

In order to further clarify the structure of state space, we numerically
determined the basins of attraction of the three attractors displayed
in Fig.\ 2a in two-dimensional sections of state space. The example
shown in Fig.\ 2b reveals that attractors are surrounded by basins
of attraction of other attractors as predicted by the above analysis.
Because of this basin structure, noise induces repeated attractor
switching among unstable attractors. Starting from the orbit defined
by (\ref{eq:syn}) the system may switch within sets of only six periodic
orbit attractors as is apparent from the basins shown in Fig.\ 2b.
However, in larger networks (cf.\ e.g.\ Fig.\ 1a) a cluster can split
up in a combinatorial number of ways and exponentially many periodic
orbit attractors are present among which the system may switch. The
larger such networks are, the higher the flexibility they exhibit
in visiting different attractors and exploring state space.

The preceeding analysis demonstrates the existence of unstable attractors.
To answer the question, how common unstable attractors actually are,
we numerically estimated the fraction \( p_{\mathrm{u}}(N) \) of
state space occupied by basins of unstable attractors. As an example,
Fig.\ 3 (inset) displays \( p_{\mathrm{u}}(N) \) for \( \varepsilon =0.2 \)
and \( \tau =0.15 \). 
\begin{figure}
{\centering {\small \resizebox*{7cm}{!}{\includegraphics{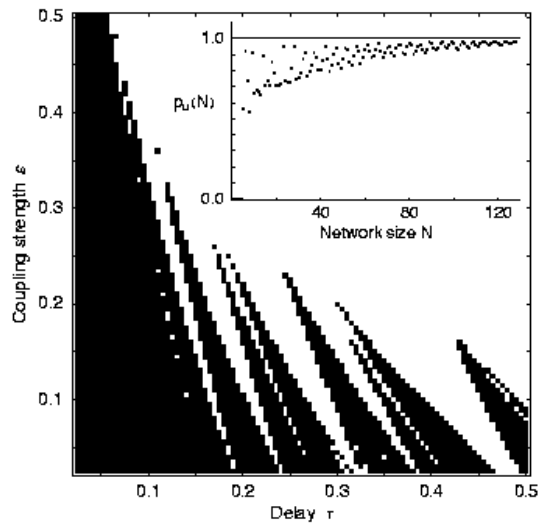}} } \par}

\caption{Unstable attractors prevail for large networks and persist in a wide
region of parameter space. Inset: \protect\( p_{\mathrm{u}}(N)\protect \)
for \protect\( N\leq 128\protect \), \protect\( \eps =0.2\protect \),
\protect\( \tau =0.15\protect \). Main figure: Parameters with \protect\( p_{\mathrm{u}}(100)>0.5\protect \)
are marked in black.}
\end{figure}
 While unstable attractors are absent if networks are too small (here
\( N\leq 4 \)) and coexist with stable attractors in larger networks,
the fraction \( p_{\mathrm{u}}(N) \) approaches one for \( N\gg 1 \).
More generally, we observed that \( p_{\mathrm{u}}(N) \) approaches
either zero or one in large networks, depending on the parameters.
For networks of \( N=100 \) oscillators Fig.\ 3 shows the region
of parameter space in which unstable attractors prevail (\( p_{\mathrm{u}}(100)>0.5 \)).
As this region covers a substantial part of parameter space, precise
parameter tuning is not needed to obtain unstable attractors. Furthermore
we find the same qualitative behavior independent of the detailed
form of \( U \). Hence, the occurrence of unstable attractors is
a robust collective phenomenon in this model class of networks of
excitatorily pulse-coupled oscillators.

Unstable attractors persist under various classes of structural perturbations.
For instance, preliminary studies on networks with randomly diluted
connectivity suggest, that a symmetric, all-to-all coupling is not
required. Moreover, it is expected that every system obtained by a
sufficiently small structural perturbation from the one considered
here will exhibit a similar set of saddle periodic orbits, because
linearly unstable states can generally not be stabilized by such a
perturbation. Although, in general, these orbits may no longer be
attracting, their dynamical consequences are expected to persist.
In particular, a switching along heteroclinic connections may occur
in the presence of noisy or deterministic, time-varying signals. As
in the original system, the sequence of states reached may be determined
by the directions into which such a signal guides the trajectory.
By increasing and decreasing the strength of this signal, the time-scale
of switching may be decreased and increased, respectively, due to
the linear instability. Interestingly, it has recently been shown
that certain models of neural networks are capable of dynamically
encoding information as trajectories near heteroclinic connections
\cite{Rabinovich}. 

Furthermore, switching among unstable states does also occur in systems
of continuously, phase-coupled oscillators \cite{Hansel,Kori} that
can be obtained from pulse-coupled oscillators in a certain limit
of weak coupling \cite{Kuramoto}. In particular, Hansel, Mato, and
Meunier show that a system of phase-coupled oscillators may switch
back and forth among pairs of two-cluster states \cite{Hansel}. Working
in the limit of infinitely fast response, i.e.\ discontinuous phase
jumps, we have demonstrated that far more complicated switching transitions
can occur in large networks if the oscillators are pulse-coupled.

In this Letter, we have presented the first example of a dynamical
system, a network of pulse-coupled oscillators, that naturally exhibits
attracting periodic orbits that are unstable. Intriguingly, these
unstable attractors are located remote from the volume of their own
basin of attraction. We have shown that they prevail in large networks
and for a wide range of parameters. Whereas unstable periodic orbits
are essential for the dynamics of many nonlinear systems, unstable
\textit{attracting} periodic orbits previously seemed to be exceptional
cases. Our results indicate that in a class of systems of pulse-coupled
units unstable attractors are the rule rather than the exception. 

We thank A.\ Aertsen, M.\ Diesmann, U.\ Ernst, D.\ Hansel, K.\ Kaneko,
K.\ Pawelzik and C.\ v.\ Vreeswijk for useful discussions.


\begin{thebibliography}{10}
\bibitem{parameters}We fix \( \eps =0.2 \), \( \tau =0.15 \) and \( U(\phi )=\ln \left( 1+(e^{b}-1)\phi \right) /b \)
with \( b=3.0 \). The orbits considered are structurally stable in
a neighborhood of these parameters and the function \( U \).
\bibitem{modeldetails}Here \( C'=C_{1}\approx 0.756\neq C=C_{0} \) because after a general
perturbation \( \boldsymbol {\delta } \) clusters are split up such
that in particular \( \phi _{1}<\phi _{2} \). Hence, oscillator \( 6 \)
receives the later signal (from osc. 1) only after it is reset by
the earlier signal (from osc. 2). Thus, the system jumps away from
the original orbit and the orbit approached in the limit \( \boldsymbol {\delta }\rightarrow \mathbf {0} \)
is different from the original one. There exist, however, other unstable
attracting periodic orbits that do not show this effect. For details
see \cite{Timme_inprep}.
\bibitem{F_i}\( \boldsymbol {F} \) is defined by \( F_{2}(\boldsymbol {\delta })=0 \),
\( F_{3}(\boldsymbol {\delta })=\Delta L+H(\tau -\Delta L)-A \),
\( F_{4}(\boldsymbol {\delta })=H(\tau +\Delta L)-A \), \( F_{5}(\boldsymbol {\delta })=H(H(1+2\tau -L_{4})+\Delta L)-B \),
\( F_{6}(\boldsymbol {\delta })=H(\Delta L+H(\tau +1-L_{3}+H(2\tau -\delta _{2}+H(\delta _{2}))))-C' \)
where we abbreviate \( H(\phi )=U^{-1}(U(\phi )+\epshat ) \), \( L_{i}\equiv L_{i}(\boldsymbol {\delta })=H(\tau +H(\tau -\delta _{2}+H(\delta _{2}+H(\tau +\delta _{i}-\delta _{2}+H(\tau ))))) \)
for \( i\in \{3,4\} \) and \( \Delta L=L_{4}-L_{3} \).
\end{thebibliography}
\end{document}